\documentclass{optica-article}

\journal{oe}
\articletype{Research Article}
\usepackage{lineno}
\usepackage{lipsum}
\usepackage{hyperref}
\usepackage{physics}
\usepackage{mathtools}
\hypersetup{
    unicode=true, % non-Latin characters in Acrobat?s bookmarks
    plainpages=false,
    colorlinks=true,% false: boxed links; true: colored links
    linkcolor=blue,% color of internal links
    citecolor=blue,% color of links to bibliography
    filecolor=blue,% color of file links
    urlcolor=blue% color of external links
}
\begin{document}		

\title{Cooperative optical pattern formation in an ultrathin atomic layer}

\author{C. D. Parmee and J. Ruostekoski\authormark{*}}

\address{Department of Physics, Lancaster University, Lancaster, LA1 4YB, United Kingdom}
\email{\authormark{*}j.ruostekoski@lancaster.ac.uk}

\begin{abstract}
Spontaneous pattern formation from a uniform state is a widely studied nonlinear optical phenomenon that shares similarities with non-equilibrium pattern formation in other scientific domains. Here we show how a single layer of atoms in an array can undergo nonlinear amplification of fluctuations, leading to the formation of intricate optical patterns. The origin of the patterns is intrinsically cooperative, eliminating the necessity of mirrors or cavities, although introduction of a mirror in the vicinity of the atoms significantly modifies the scattering profiles. The emergence of these optical patterns is tied to a bistable collective response, which can be qualitatively described by a long-wavelength approximation, similar to a nonlinear Schr\"odinger equation of optical Kerr media or ring cavities. These collective excitations have the ability to form singular defects and unveil atomic position fluctuations through wave-like distortions.
\end{abstract}

\date{\today} 

\section{Introduction}
Pattern formation is ubiquitous throughout different fields of science, emerging from dynamical instabilities in nonlinear systems~\cite{Cross1993}. 
In chemistry and biology, there are numerous examples of pattern formation~\cite{Meinhardt1992, Maini1997}, such as in precipitation reactions with the formation of Liesegang rings~\cite{Nabika2020}, and in diffusion reactions leading to Turing patterns~\cite{Turing1952} which can explain markings on animal fur and shells. 
In physics, the formation of patterns has been studied in hydrodynamics with, e.g., Rayleigh-B\'enard convection and Taylor-Couette flow~\cite{Fauve1998},
and in liquid crystals~\cite{Macdonald1992}.
Significant focus has been given to pattern formation in nonlinear optics~\cite{Lugiato1987,Grynberg1988,Firth1990,DAlessandro1991,HAELTERMAN1992,Ackemann95,Afanasev1995,Loiko1996,Arecchi1999,Schapers00,Castelli2017},
where small fluctuations in the light field are amplified due to a nonlinear medium,
either in a ring cavity or with feedback from a single mirror. Pattern formation in nonlinear optics has been studied more recently in systems of driven atom clouds in single~\cite{Scroggie94,Domokos2002,Black03,Asboth2005,Lee14,Baumann2010,Caballero-Benitez2015} and multi-mode~\cite{Vaidya18} cavities, and in atom clouds with optomechanical back-action~\cite{Labeyrie2014,Baio21}.

The formation of patterns in nonlinear optics is typically described by the dynamics of the dissipative nonlinear Schr\"odinger equation (NLSE)~\cite{Castelli2017}
\begin{align}\label{Eq:DNLSELight}
	\text{i}\frac{\partial\psi}{\partial\xi}&=(\theta-\text{i}) \psi + \text{i}y - A|\psi|^2\psi - \bar{\nabla}_{\perp}^2 \psi,
\end{align}
with a field $\psi$ that can form patterns under a driving field $y$, phase rotation $\theta$, and with a Kerr nonlinear magnitude $A$. 
The dynamics of many physical configurations can be mapped to Eq.~\eqref{Eq:DNLSELight} under certain limits.
For light in a ring cavity, $\psi$ and $y$ are the internal and external electric fields, respectively, $\theta$ the light-cavity detuning, and $A=\pm 1$ for light focusing or defocusing.
In ring cavities, $\xi$ represents the dimensionless time coordinate $\xi=\alpha t$, where $\alpha$ denotes the characteristic time scale of the system given by the cavity decay rate $\kappa$.
However, in other systems $\xi$ can also be a dimensionless spatial coordinate, i.e., $\xi=\alpha z$, in the slowly-varying envelope approximation of the propagating wave~\cite{meystre1998} with, e.g., $\alpha=k$ where $k$ is the wavenumber of light. Sampling the pattern along $\xi$ is then equivalent to taking snapshots of patterns evolving in time.
In the latter case, the spatial derivative acts as a kinetic energy term for the dynamics perpendicular to the direction of light propagation, where, similar to the time derivative, spatial derivatives are with respect to dimensionless spatial coordinates $\bar{\bf{r}} = \beta \bf{r}$. Here, $\beta$ is a characteristic length scale of the system, e.g., in the ring cavity $\beta = T/\sqrt{\lambda \mathcal{L}}$ with cavity length $\mathcal{L}$ and transmission coefficient of the mirrors $T$. 

In classic studies of optical pattern formation in a ring cavity, Eq.~\eqref{Eq:DNLSELight} is also known as the Lugiato-Lefever equation (LLE)~\cite{Lugiato1987} where the electric field forms patterns due to the interplay between diffraction and the Kerr nonlinearity.
The onset of pattern formation predominantly occurs for parameters where Eq.~\eqref{Eq:DNLSELight} exhibits bistability, where there are two stable solutions for the field amplitude in the cavity. However, patterns can also emerge when there is only a single stable solution.
The Kerr nonlinearity results in four-wave mode mixing where the medium absorbs two incident field photons with transverse wavevector ${\bf k}= 0$ and emits two with $\pm {\bf k}$. This leads to a stripe pattern, while further four-wave mixing of these photons results in more complex structures, such as hexagonal patterns. As well as global patterns, localized soliton solutions can also form~\cite{Odent2011,Minardi10,Firth96,Azam2022}.

Currently, there is rapid development in the experiments on the optical properties of planar arrays of regularly spaced atoms confined in optical lattices~\cite{Rui2020,Srakaew22}. 
Due to the subwavelength spacing and negligible motion, the atoms are coupled to resonant light and exhibit strong collective behavior. Such ultrathin, single atomic layer surfaces are therefore of great interest as they can mimic~\cite{Jenkins2012a,Ruostekoski23} the optical behavior of artificially fabricated metasurfaces~\cite{Yu14}. 
Furthermore, recent studies have shown that atomic arrays now offer possibilities for realizing strong cooperative nonlinear optical responses~\cite{Parmee2018,williamson2020b,Cidrim20,Parmee2020,Parmee2021,Bettles2020,Zhang2022,Ferioli21,Holzinger21,Rusconi2021,Moreno2021,Parmee22b,Pedersen23,Rubies-Bigorda23,Robicheaux23}.

Here, we show how patterns emerge in the optical excitations in regular arrays of cold atoms. We simulate the dynamics of atoms with a two-level and an isotropic transition, both with and without a mirror placed near the array. 
Many early examples of optical pattern formation involve feedback between a sample and a mirror~\cite{Lugiato1987,Firth1990,DAlessandro1991,Arecchi1999}, where weak fluctuations in a nonlinear medium lead to scattering of light in a direction that depends on the excitation wavevector. This scattered light reflects off the mirror and back to the sample, amplifying the initial excitation to recognizable patterns at sufficient intensities and frequencies of the incident light, typically in regimes where optical bistability occurs. This mechanism of pattern formation has also been analyzed for nonlinear layers~\cite{Afanasev1995,Loiko1996}.
Here, however, we find that the optical response of cold atoms in an array alone is sufficient to amplify fluctuations.
This is possible because of the strong cooperative resonant dipole-dipole interactions between closely spaced atoms that are capable of generating nonlinear couplings between atomic excitations and \emph{intrinsic} optical bistabilities~\cite{Parmee2021}, eliminating the necessity of feedback mechanisms through mirrors or cavities.

The amplification of fluctuations within the ensemble, driven by its nonlinearity, gives rise to the development of spatially varying atomic coherence. This phenomenon not only leads to the emergence of intricate patterns but also causes a shift in the direction of the scattered light, which conveys information about the underlying amplified atomic excitations. We find that introducing a mirror near the array further amplifies certain modes by allowing the scattered light to reflect back to the atoms, leading to more exotic patterns than found in the mirrorless case, and also rotation of the atomic dipoles in arrays of atoms with an isotropic transition.
We analyze the robustness of the patterns to decreasing atom density and demonstrate how the patterns can persist in the presence of spatial disorder. Our findings reveal that the position fluctuations due to finite atom confinement lead to wave-like pattern distortions or the loss of patterns at specific parameter values, while increasing the lattice constant results in the loss of patterns consistently with the previous analysis of the bistability thresholds~\cite{Parmee2021}. 
Furthermore, we introduce singular defects in the excitations by vortex beams, which allows us to study the emergence of nonlinear defects that are observable in the profiles of the phase and polarization ellipses. 

Our analysis demonstrates how a single ultrathin atomic layer exhibits pattern formation analogous to nonlinear Kerr media placed next to a mirror or inside cavities, and described by a dissipative NLSE or LLE. Inspired by this connection, we construct for the collective atom response a qualitative long-wavelength approximation that maps the dynamics to a NLSE. The NLSE demonstrates both the onset of pattern formation in regimes of optical bistability, and produces patterns with comparable geometry to those found in the full dynamics. Furthermore, the dynamical analogy between strong light-mediated interactions in dense atomic arrays and nonlinear optics of Kerr media or atomic superfluids points towards other exciting possibilities of soliton and vortex formation.

%%%%%%%%%%%%%%%%%%%%%%%%%%%%%%%%%%%%%%%%%%%%%%%%%%%%%%%%%%%%%%%%%%%%%%%%%%%%
\section{Model}\label{sec:setup}

We study a single layer of atoms trapped in a two-dimensional (2D) square array within the $xy$ plane, illuminated by a near monochromatic incident field, with the positive frequency component $ \boldsymbol{\mathcal{E}}{}^+(\textbf{r})=\boldsymbol{\mathcal{E}}_0e^{\text{i}kz}$, 
wavevector $\textbf{k}$, frequency $\omega =ck$, and where $\boldsymbol{\mathcal{E}}_0$ is either a spatially uniform beam with polarization $(\hat{{\bf x}}+\hat{{\bf y}})/\sqrt{2}$ or a superposition of Laguerre-Gaussian beams with the polarization in the $xy$ plane.   
The incident field drives a two-level or an isotropic multilevel $\ket{J=0, m_J=0}\rightarrow\ket{J' = 1, m_{J'}=\mu}$ atomic transition, inducing an atomic polarization density operator with positive frequency component,
\begin{equation}
	\hat{\textbf{P}}{}^+(\textbf{r})=\mathcal{D}\sum_{j, \mu}\delta(\textbf{r}-\textbf{r}_j)\hat{\textbf{e}}_{\mu} \hat{\sigma}_{j\mu}^{-},
\end{equation}
where $\mathcal{D}$ is the reduced dipole matrix element, $\ket{g}_j$ the ground state, $\ket{\mu}_j$ the excited states and $\hat{\sigma}_{j\mu}^{-}=|g\rangle_{j}\mbox{}_{j}\langle \mu|=(\hat{\sigma}_{j\mu}^+)^{\dagger}$.
The Hamiltonian of the atomic interaction terms with light is obtained by making a rotating wave approximation to remove the fast co-rotating terms, giving~\cite{CohenT}
\begin{equation}\label{Eq:Hamiltonian}
\hat{H}=-\hbar\Delta\sum_{j,\mu}\hat{\sigma}^{\mu \mu}_j
-\frac{1}{\epsilon_0}\int d^3\textbf{r} \left( \hat{\textbf{D}}{}^+(\textbf{r})\cdot \hat{\textbf{P}}{}^-(\textbf{r}) + {\rm H.c.}\right),
\end{equation} 
where $\Delta=\omega - \omega_{0}$ denotes the  laser frequency detuning  from the atomic resonance frequency $\omega_0$ and $\hat{\sigma}^{\mu \mu}_j=\hat{\sigma}_{j\mu}^{+}\hat{\sigma}_{j\mu}^{-}$. The positive frequency component of the electric displacement operator is represented by $ \hat{\textbf{D}}{}^+$.
The Hamiltonian also contains the self-polarization term
that is inconsequential in our system of nonoverlapping point atoms~\cite{Ruostekoski1997a,Lee16}. 
In addition to the atom-light coupling terms of Eq.~\eqref{Eq:Hamiltonian}, the free electromagnetic field energy is represented by the standard quantized expression~\cite{CohenT}. 
Observables are expressed in terms of slowly varying field amplitudes and atomic variables, $\hat{\textbf{D}}{}^+e^{i\omega t} \rightarrow \hat{\textbf{D}}{}^+$ and $\hat{\sigma}{}^{-}_{j \mu}e^{i\omega t} \rightarrow \hat{\sigma}{}^{-}_{j \mu}$, and the incident light is expressed through the Rabi frequencies ${\cal R}_{g\mu}^{(j)}=\mathcal{D}\hat{\textbf{e}}_{\mu}^*\cdot\boldsymbol{\mathcal{E}}{}^+(\textbf{r}_j)/\hbar$, with corresponding intensities $I^{(j)}/I_{\rm sat}=2\sum_{\mu}|{\cal R}_{g\mu}^{(j)}/\gamma|^2$ and saturation intensity $I_{\rm sat}= 4\pi^2\hbar c \gamma/3\lambda^3$. 

\begin{figure}
	\hspace*{-0cm}
	\center
	\includegraphics[width=0.9\columnwidth]{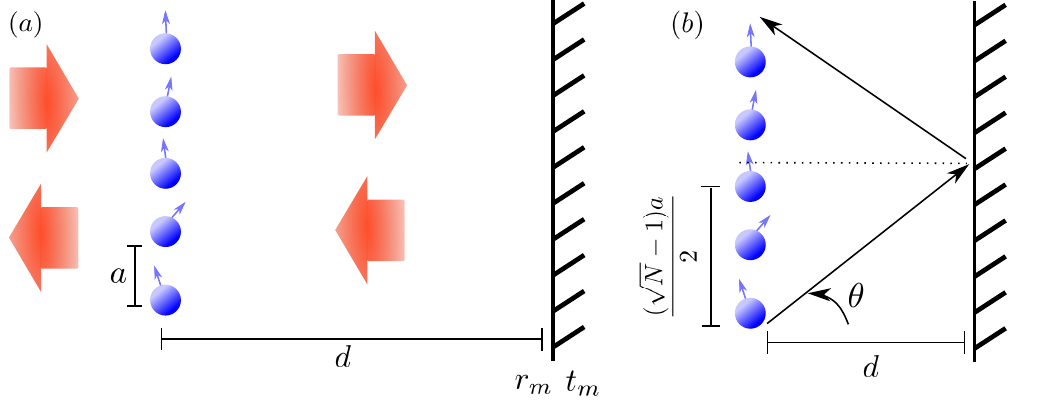}
	\vspace{-0 cm}
	\caption{A planar array of atoms driven by incident light. (a) The atoms are trapped in a regular square array with spacing $a$ and at a distance $d$ from a mirror with reflection amplitude $r_m$.
 (b) Excitations in the array emit light at an angle $\theta$ which reflects off the mirror and back to the array, causing the formation of patterns.
	}
	\label{Fig:Model}
\end{figure}

The total field at each atom 
$\hat{\textbf{E}}{}^+(\textbf{r}) =  \boldsymbol{\mathcal{E}}{}^+(\textbf{r})+\hat{\textbf{E}}{}^+_s(\textbf{r})$ with scattered light 
\begin{equation} \label{Efieldscattered}
	\epsilon_0\hat{\textbf{E}}{}^+_{s}(\textbf{r}) =\sum_{j,\mu}\mathsf{G}(\textbf{r}-\textbf{r}_j)\mathcal{D}\hat{\textbf{e}}_{\mu} \hat{\sigma}_{j\mu}^{-},
\end{equation}
where the dipole radiation kernel acting on a dipole located at the origin, with $r= |\textbf{r}|$ and $\hat{\mathbf{r}}=\textbf{r}/r$, is given by~\cite{Jackson}
\begin{align}\label{Gdef}
	\mathsf{G}(\mathbf{r})\mathbf{d}&=-\frac{\mathbf{d}\delta(\mathbf{r})}{3}+\frac{k^3}{4\pi}\Bigg\{\left(\hat{\mathbf{r}}\times\mathbf{d}\right)\times\hat{\mathbf{r}}\frac{e^{\text{i}kr}}{kr}-\left[3\hat{\mathbf{r}}\left(\hat{\mathbf{r}}\cdot\mathbf{d}\right)-\mathbf{d}\right]\left[\frac{\text{i}}{(kr)^2}-\frac{1}{(kr)^3}\right]e^{\text{i}kr}\Bigg\}.
\end{align}
In a tightly-confined array, the atoms are fixed at the coordinates defined by the lattice. In typical experimental situations the atoms are trapped by a periodic optical lattice with one atom per site and undergo position fluctuations in the vibrational ground states of the individual sites~\cite{Rui2020,Srakaew22}. We incorporate the effects of fluctuating positions by stochastically sampling the positions $\{{\bf r}_1,\ldots,{\bf r}_N\}$ of $N$ atoms using a Gaussian density distribution with root-mean-square width $\sigma$ at each site~\cite{Jenkins2012a}.
The coupled dynamics between the atoms and light is simulated within a semiclassical approximation by ignoring quantum fluctuations between different atoms~\cite{Lee16,Bettles2020}. The system dynamics for any particular stochastic configuration of fixed atomic positions is then determined by the following nonlinear equations
\begin{subequations}\label{Eq:SpinEquations}
	\begin{align}
		\dot{\rho}^{(j)}_{g \mu}=&[\text{i}\Delta-\gamma]\rho^{(j)}_{g \mu}+\text{i}\tilde{\mathcal{R}}_{\mu}^{(j)}\rho_{gg}^{(j)}-\text{i}\sum_{\alpha}\tilde{\mathcal{R}}_{\alpha}^{(j)}\rho_{\alpha\mu}^{(j)},\\
		\dot{\rho}^{(j)}_{\mu \nu}=&-2\gamma{\rho}^{(j)}_{\mu \nu}+\text{i}\tilde{\mathcal{R}}_{\nu}^{(j)}(\rho_{g\mu}^{(j)})^* -\text{i}(\tilde{\mathcal{R}}_{\mu}^{(j)})^*\rho_{g\nu}^{(j)},
	\end{align}
\end{subequations}
where $\gamma = \mathcal{D}^2 k^3/(6\pi\epsilon_0\hbar)$ is the single atom linewidth.
The terms $\rho_{\mu \nu}^{(j)}=\langle\hat{\sigma}{}^{\mu\nu}_{j}\rangle$, where $\hat{\sigma}_{j}^{\mu\nu}=|\mu\rangle_{j}\mbox{}_{j}\langle \nu|$, represent excited state populations ($\mu=\nu$), and coherences between different excited states ($\mu\neq\nu$), while $\rho_{g \mu}^{(j)}=\langle\hat{\sigma}{}^{-}_{j\mu}\rangle$ are coherences between the ground and excited states. The effective Rabi frequencies $\tilde{\mathcal{R}}_{\nu}^{(j)}$ include the contributions from the incident field driving the atom $j$ and from the scattered fields from all the other atoms in the system~\cite{Parmee2020}. The dipole-dipole interactions resulting from the multiple scattering between the atoms give rise to a cooperative response that is sensitive to the atomic positions. In the absence of a collective response, the effective Rabi frequencies $\tilde{\mathcal{R}}_{\nu}^{(j)}$ in Eq.~\eqref{Eq:SpinEquations} are replaced by the single-atom Rabi frequencies ${\mathcal{R}}_{\nu}^{(j)}$, resulting in the standard optical Bloch equations. Equations~\eqref{Eq:SpinEquations} form a complete set where the ground state populations are obtained through the population conservation 
$\rho_{gg}^{(j)}= 1- \sum_{\mu}\rho_{\mu \mu}^{(j)}$.

We consider two configurations in this paper: the array without a mirror and the array with a mirror placed a distance $d$ from the array (Fig.~\ref{Fig:Model}) with reflection amplitude $r^{(\mu)}_m$, depending on the polarization $\mu$ with the quantization axis along the $z$-direction. It is assumed that the time taken for light to travel to and from the mirror is negligible. 
In the presence of a mirror, the effective Rabi frequencies are
\begin{align}\label{Eq:EffectiveRabi}
	\tilde{\mathcal{R}}_{\mu}^{(j)} &= \mathcal{R}_{g\mu}^{(j)}(1-r_{m}e^{2\text{i}k d}) +\frac{6\pi\gamma}{k^3}\hat{\textbf{e}}_{\mu}^*\cdot \left[  \sum_{l \neq j,\nu}^{} \mathsf{G}(\textbf{r}_j-\textbf{r}_l)\hat{\textbf{e}}_{ \nu} \rho_{g\nu}^{(l)}+ \sum_{l,\nu}^{} r^{(\nu)}_{m} \mathsf{G}(\textbf{r}_j-\textbf{r}_l')\hat{\textbf{e}}_{ \nu} \rho_{g\nu}^{(l)} \right].
\end{align}  
The first term in Eq.~\eqref{Eq:EffectiveRabi} is the incident field and its reflected counterpart for which $\mathcal{R}_{g0}^{(j)}=0$, while the second is the scattered field from all other atoms at positions ${\bf r}_l$ in the ensemble. The third term is the scattered field from the image atoms located at ${\bf r}_l'$ due to the mirror, where we have the polarization reflection coefficients $ r^{(\nu)}_{m} = -r_m$ ($ r^{(\nu)}_{m} = r_m$) for $\nu=\pm1$ ($\nu = 0$). The case without a mirror is obtained by setting $r_{m}=0$. 
The scattered fields drive the atom at ${\bf r}_j$, and describe light-mediated dipole-dipole interactions through $\mathsf{G}$, leading to recurrent light scattering between the atoms. 
The third term in Eq.~\eqref{Eq:EffectiveRabi} is responsible for coupling the polarization components $ \rho_{g \nu}^{(l)} $, with $\nu=\pm1$, to excitations $ \rho_{g 0}^{(j)} $.

A two-level transition can be obtained by tuning all but one excited level off-resonance, e.g., by magnetic fields or in alkali-metal atoms by using a cycling transition with additional levels off-resonance. The dynamics then corresponds to a collection of driven two-level atoms
\begin{subequations}\label{Eq:SpinEquations2LS}
\begin{align}
	\dot{\rho}^{(j)}_{ge}=&\left(\text{i}\Delta-\gamma\right) {\rho}^{(j)}_{ge}-\text{i}(2{\rho}^{(j)}_{ee}-1)\tilde{\mathcal{R}}_{e}^{(j)},\label{Eq:Coherence}\\
	\dot{\rho}^{(j)}_{ee}=&-2\gamma {\rho}^{(j)}_{ee}+2\text{Im}[(\tilde{\mathcal{R}}_{e}^{(j)})^*{\rho}^{(j)}_{ge}],\label{Eq:Excitations}
\end{align}
\end{subequations}
where we have assumed the atomic dipole is oriented along the incident light polarization.
We focus mainly on the dynamics of Eqs.~\eqref{Eq:SpinEquations2LS} throughout the paper, but also consider the multilevel isotropic $J=0\rightarrow J'=1$ transition.

%%%%%%%%%%%%%%%%%%%%%%%%%%%%%%%%%%%%%%%%%%%%%%%%%%%%%%%%%%%%%%%%%%%%%%%%%%%%%%%%%%%%%%%%

\subsection{Patterns with a two-level transition}\label{Sec:2LS}

\begin{figure}
	\hspace*{-0cm}
	\includegraphics[width=\columnwidth]{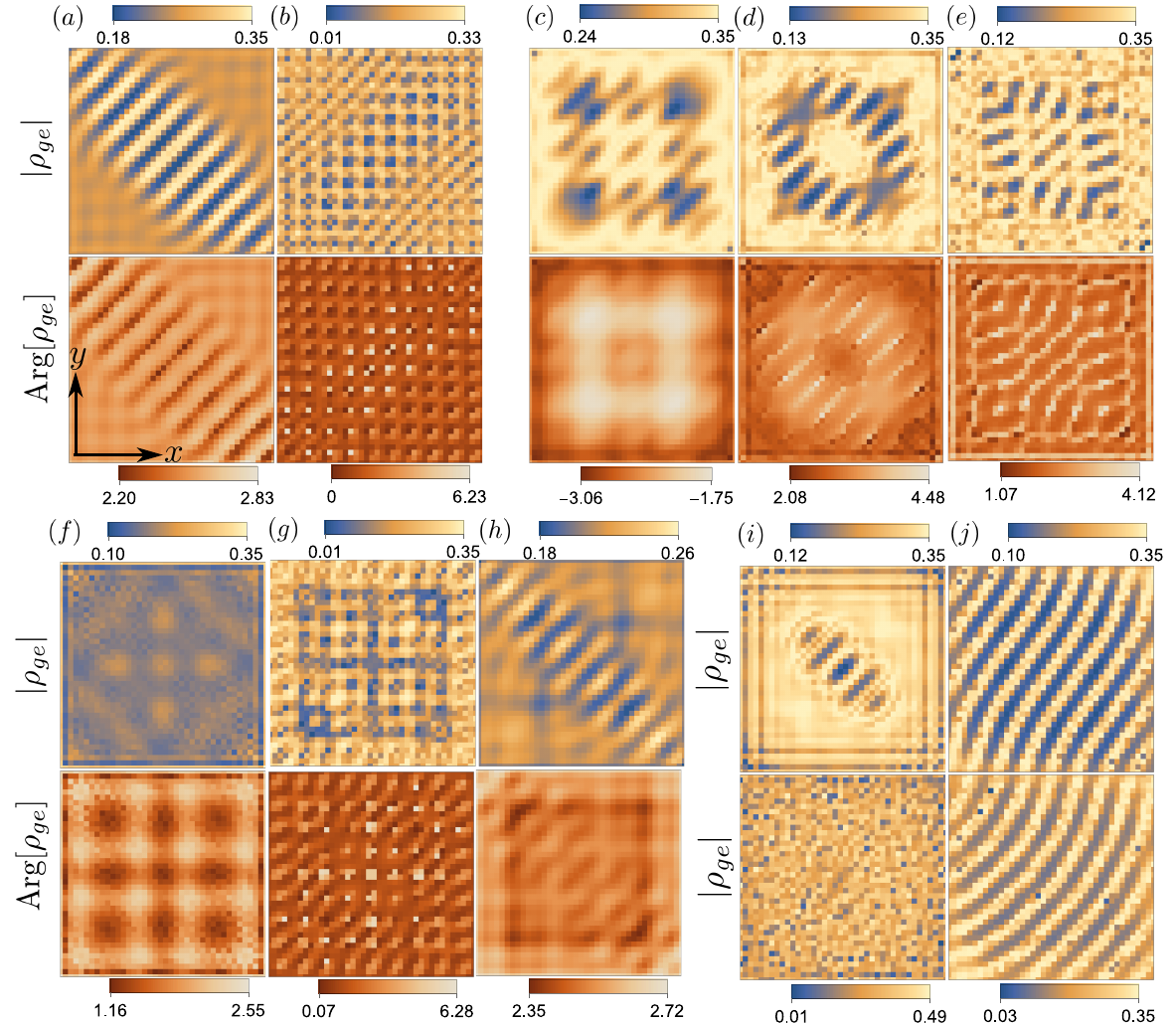}
	\vspace{-0cm}
	\caption{Pattern formation in the atomic coherence $\rho_{ge}$ for an $N=1600$ atomic array with $a=0.2\lambda$ driven by a plane wave. 
        (a, b) With no mirror present, (a) stripes can form for an incident field intensity $I/I_{\rm sat}=50$ and laser frequency detuning from the atomic resonance $\Delta/\gamma = 2.5$ and (b) a checkerboard pattern forms for $I/I_{\rm sat}=2$ and $\Delta/\gamma=-2$. 
		With a mirror at (c,d,e) $d=2.48\lambda$, more exotic patterns emerge when (c) $I/I_{\rm sat}=98$ and $\Delta/\gamma = 2$, (d) $I/I_{\rm sat}=50$ and $\Delta/\gamma = 0.5$, and (e) $I/I_{\rm sat}=50$ and $\Delta/\gamma = -0.5$.
		Patterns for a mirror at (f,g,h) $d=2.28\lambda$, when (f) $I/I_{\rm sat}=2$ and $\Delta/\gamma = 1.5$, (g) $I/I_{\rm sat}=2$ and $\Delta/\gamma = -3.5$, and (h) $I/I_{\rm sat}=18$ and $\Delta/\gamma = 2$.
		(i,j) Patterns for atoms without (top panels) and with (bottom panels) position fluctuations with standard deviation $\sigma = 0.1a$, where the pattern at (i) $I/I_{\rm sat}=18$ and $\Delta/\gamma = 0$ is lost or, at (j) $I/I_{\rm sat}=162$ and $\Delta/\gamma = -0.5$, distorted due to the disorder.
	}
	\label{fig:patterns}
\end{figure}

We first consider the dynamics of two-level atoms in a large square array driven by a spatially uniform incident field without a mirror.
Evolving Eqs.~\eqref{Eq:SpinEquations2LS} in time for various frequencies and intensities of the incident field, we find many patterns emerging in the steady-state atomic coherence, e.g., a striped pattern shown in Fig.~\ref{fig:patterns}(a) or checkerboard pattern in Fig.~\ref{fig:patterns}(b). 
Despite a spatially-uniform incident field profile, dipole-dipole interactions lead to spatially-nonuniform configurations, where the pattern orientation is determined by the incident field polarization. 
The discrete Fourier transform of the atomic coherence shows strong peaks usually at only one or two wavevectors ${\bf q}$, which indicates the striped and checkerboard patterns are formed from only one or two atomic excitation modes. Because we have a large array, the wavevectors of excited modes are reasonably well-defined, with lattice edge effects resulting in modes approximately described by sine and cosine waves with $\pm {\bf q}$ pairs forming. 
Varying the incident field frequency and intensity, we find several variations of the striped or checkerboard patterns. Most of these variations are still composed of one or two modes, but some show small peaks at other wavevectors in the discrete Fourier transform, indicating that several other modes are weakly excited due to the nonlinear interactions.

\begin{figure}
	\hspace*{-0cm}
	\center
	\includegraphics[width=\columnwidth]{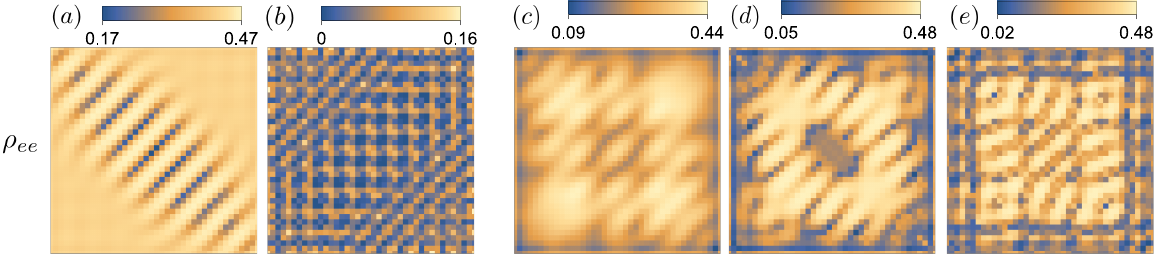}
	\vspace{-0 cm}
	\caption{Changes in the excited state population $\rho_{ee}$ when patterns form, with (a,b,c,d,e) corresponding to the configurations and coherences $\rho_{ge}$ in Fig.~\ref{fig:patterns}(a,b,c,d,e), respectively. The patterns in the $\rho_{ee}$ are qualitatively similar to those found in $\rho_{ge}$.
	}
	\label{Fig:ExcitedStatePopulation}
\end{figure}

The excited modes can be determined by the scattered light which depends on ${\bf q}$. At a point ${\bf r} = (x,y,z)$ sufficiently far from the array,
but still separated less than the array size, we have for the coherently scattered field amplitude~\cite{Javanainen2019} 
\begin{align}\label{Eq:FF_radiation}
\langle \hat{{\bf E}}_s^+({\bf r})\rangle = \frac{\text{i}\mathcal{D}}{2\mathcal{A}\epsilon_0}P({\bf q})\,\hat{{\bf e}}\,\rho_{ge}e^{\text{i}(q_x x + q_y y + \sqrt{k^2-q_x^2-q_y^2}|z|)}.
\end{align}
In Eq.~\eqref{Eq:FF_radiation}, $P_{\mu \nu}({\bf q}) = (k^2\delta_{\mu \nu}- q_{\mu}q_{\nu})/\sqrt{k^2-q_x^2-q_y^2}$ is the projection matrix to the subspace orthogonal to the light propagation, $\mathcal{A}$ the area of a unit cell of the lattice, and we have used the fact an atomic excitation with wavevector ${\bf q}$ in an infinite array has a coherence $\rho_{ge}^{(j)} = \rho_{ge}e^{\text{i}{\bf q}\cdot {\bf r}_j}$.
Emission from the excitation is a plane wave which propagates along the direction $\hat{\bf k}=[\sin\theta \cos\phi, \sin\theta \sin\phi, \cos\theta]$, and therefore the angle of emission of the scattered light from a mode with wavevector ${\bf q}$ is given by $|{\bf q}|/k = \sin \theta$. The coherently scattered light could be detected by homodyne measurements, while the scattering direction reveals the excitation wavevectors.
The patterns emerging in the coherence also have corresponding changes in the excited state population, e.g., Fig.~\ref{Fig:ExcitedStatePopulation}, which would result in observable signatures of the patterns in the incoherently scattered light.

We find that the patterns in the atomic polarization and in the scattered light emerge solely due to the light-mediated cooperative interactions between the atoms, which is in contrast to the traditional optical studies without cooperative coupling, where patterns can only form by the feedback of scattered light from the sample and the mirror~\cite{Lugiato1987,DAlessandro1991,HAELTERMAN1992,Domokos2002,Black03,Asboth2005,Lee14,Baumann2010,Caballero-Benitez2015}.
In such systems, the mechanism of pattern formation is through four-wave mixing where weak excitations in the medium scatter the incident light in a direction that depends on the excitation wavevector ${\bf q}$. 
This scattered light reflects off the mirror and back to the sample, changing phase in the process and amplifying the initial excitation. 
However, while a mirror is not needed for the formation of patterns in our system, we find introducing a perfectly reflecting one near the array results in more exotic patterns emerging [Fig.~\ref{fig:patterns}(b,c)] due to the scattering of light off the mirror from excited modes. The new patterns now exhibit longer spatial variation and four-fold symmetry, with the Fourier transform of the atomic coherence showing several peaks indicating that many modes are now excited due to the nonlinear interactions, and the coupling of the modes to the mirror. 

The back-action of the mirror on the atoms strongly depends on the excitation modes of the atoms due to the ${\bf q}$ dependence of the light scattering direction. In an infinite array, an excitation ${\bf q}$ perpendicular to the incident light tilts the scattering direction in a such a way that the light reflected from the mirror back to the atoms accumulates a phase slip $\exp[i 2d |{\bf q}|^2/(2k)]$~\cite{Firth1990,DAlessandro1991}.
A finite-size lattice provides additional effects. This can be understood by the geometry of the system (Fig.~\ref{Fig:Model}) where we assume that all the scattered light from the mode ${\bf q}=0$ scatters back to the atoms. For increasing ${\bf q}$, the scattered light that interacts with the atoms monotonically decreases, and not all modes will be affected by the mirror. Only modes with $|{\bf q}| < q_m$ where $q_m$ is some maximum wavevector magnitude will scatter light that interacts with atoms after reflecting back from the mirror. 
To determine $q_m$, we consider the scattered light from atoms at the bottom of the array [see Fig.~\ref{Fig:Model}(b)] which reflects off the mirror and back to the array's upper edge if $\tan \theta_m = (\sqrt{N}-1)a/2d $, while it misses the upper edge if $\theta>\theta_m$.
Substituting the maximum angle at which light no longer reflects into the expression for the light scattering angle for an excited mode $|{\bf q}|/k = \sin\theta$ gives $q_m/k = (\sqrt{N}-1) a /[(\sqrt{N}-1)^2a^2+4d^2]^{1/2}\simeq [1+4d^2/(Na^2)]^{-1/2}$ where we have used $N \gg 1$. Light from modes with $|{\bf q}|>q_m$ therefore escapes the system.
For our system, we have $q_m = 0.85k$ and so most modes in our system are enhanced by the presence of a mirror. However, even some modes with $|{\bf q}| > q_m$ may get modified by the mirror due to the cooperative interactions.

The simulations of the effect of atomic position fluctuations due to finite confinement of individual array sites lead to fluctuations in the dipole-dipole interactions between the atoms, and also alter how far an atom is from the mirror. 
We find that the disorder in the atomic positions results in wave-like distortions in the emerging patterns, e.g., in Fig.~\ref{fig:patterns}(g).
In some other cases, the disorder destroys the patterns, e.g., in Fig.~\ref{fig:patterns}(f).
Patterns are also lost for larger atom spacings both with and without a mirror near the array, where for 
%$a \gtrsim 0.2\lambda$
$a >0.2\lambda$, only small scale checkerboard patterns and some stripe phases persist. 
For 
%$a \gtrsim 0.3\lambda$
$a > 0.3\lambda$, almost all patterns are lost, and therefore sufficiently high atomic densities are needed for patterns. This is consistent with the previous results that demonstrated how the bistable behavior only emerges at sufficiently small interatomic separations~\cite{Parmee2021}, indicating that the existence of multiple stable solutions in a uniformly driven array is closely linked to pattern formation.

\begin{figure}
	\hspace*{-0cm}
	\center
	\includegraphics[width=0.9\columnwidth]{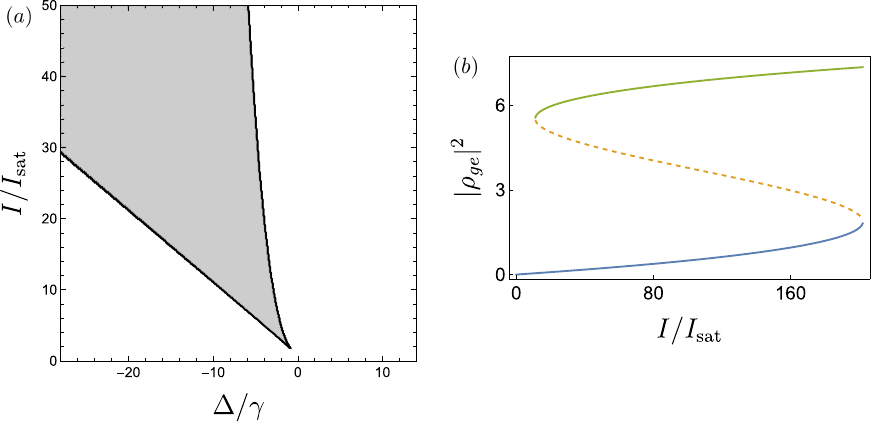}
	\vspace{-0cm}
	\caption{Bistability from the dissipative discrete nonlinear Schr\"odinger equation, Eq.~\eqref{Eq:DNLSEScaled}, for an array with $a=0.54\lambda$ with no mirror. (a) Bistable solutions (shaded region) as a function of detuning and intensity. (b) The two bistable solutions of the NLSE for sufficient intensities when the laser frequency detuning from the atomic resonance $\Delta/\gamma = -10$. The orange dashed line shows the third unstable solution.
	}
	\label{fig:bistabilityDNLSE}
\end{figure}
\begin{figure}[h]
	\hspace*{-0cm}
	\center
	\includegraphics[width=\columnwidth]{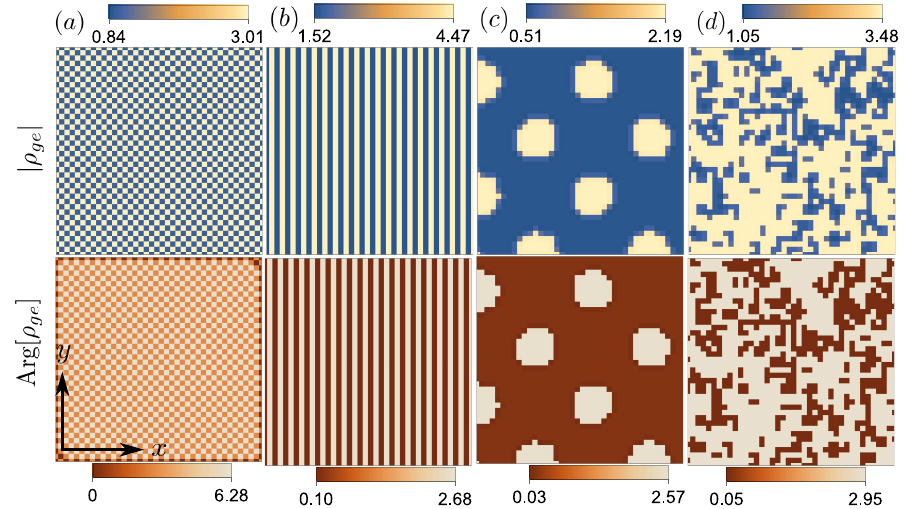}
	\vspace{-0cm}
	\caption{Emerging patterns under the dynamics of the discrete nonlinear Schr\"odinger equation, Eq.~\eqref{Eq:DNLSEScaled}, for an array with $a=0.54\lambda$ with no mirror. For parameters where bistability occurs, patterns form if seeded by changing the initial conditions. Similar patterns to the full dynamics with (a) checkerboard patterns for an intensity $I/I_{\rm sat} = 200$ and laser frequency detuning from the atomic resonance $\Delta/\gamma = -15$ and (b) striped patterns for $I/I_{\rm sat} = 338$ and $\Delta/\gamma = -10$. Other patterns not found in the full dynamics with (c) four-fold symmetry forming with appropriate initial conditions for $I/I_{\rm sat} = 32$ and $\Delta/\gamma = -7$, and (d) clumps with disordered initial conditions for $I/I_{\rm sat} = 648$ and $\Delta/\gamma = -18$.
	}
	\label{fig:patternsDNLSE}
\end{figure}

Our results illustrate how a single atomic layer exhibits pattern formation reminiscent of nonlinear Kerr media placed next to a mirror or inside cavities.
Since such systems are normally well described by a NLSE or LLE, it is illuminating to explore further connections between a collectively interacting atomic array and a NLSE. 
To find a closer link between the two representations, we simplify Eq.~\eqref{Eq:Coherence} by making the substitution $\rho_{ee}^{(j)}\approx|\rho_{ge}^{(j)}|^2$ and truncate the interatomic coupling to only include nearest-neighbor terms.
By taking the Fourier transform of the coherence
$\rho_{ge}^{(j)} = \sum_{\textbf{q}}\tilde{\rho}(\textbf{q})e^{\text{i}\textbf{q}\cdot \textbf{r}_j}$, the dipole summation terms in Eq.~\eqref{Eq:Coherence} can be rewritten
\begin{align}\label{Eq:Intermdiate3}
\sum_{j\neq l}G_{jl}\rho_{ge}^{(j)}
&\approx G_{12}\sum_{\textbf{q}}\tilde{\rho}_{ge}(\textbf{q})[4-a^2(q_x^2+q_y^2)] = 4G_{12}\rho_{ge}(\textbf{r}) - G_{12}a^2\grad^2_{\perp} \rho_{ge}(\textbf{r}),
\end{align}
where 
$G_{jl} = 6\pi\gamma/k^3 \hat{{\bf e}}\cdot\mathsf{G}(\textbf{r}_j-\textbf{r}_l)\hat{{\bf e}}$, and we have taken the long wavelength limit $|{\bf q}|a\ll 1$ to expand about the band minimum and converted back to real space to obtain the derivative term.
Eq.~\eqref{Eq:Coherence} in the continuum limit is now 
\begin{align}\label{Eq:Intermdiate4}
	\frac{\partial \rho_{ge}(\textbf{r})}{\partial t}&\simeq \left(\text{i}\Delta-\gamma+4\text{i}G_{12}\right) {\rho}_{ge}(\textbf{r})+\text{i}\mathcal{R}(\textbf{r}) - \text{i}G_{12}a^2\grad^2_{\perp} \rho_{ge}(\textbf{r})\nonumber\\
	&-2\text{i}|\rho_{ge}(\textbf{r})|^2\left[\mathcal{R}(\textbf{r})+4G_{12}\rho_{ge}(\textbf{r}) - G_{12}a^2\grad^2_{\perp}\rho_{ge}(\textbf{r})\right].
\end{align}
Retaining only the term cubic in $\rho_{ge}$ in the second line, we obtain the NLSE
\begin{align}\label{Eq:DNLSEScaled}
\text{i}\frac{\partial \rho_{ge}({\bf r})}{\partial \bar{t}}&=\left[\text{i}\left(\frac{\gamma\Omega_{12}-\Delta \Gamma_{12}}{8|G_{12}|^2}\right)+\left(\frac{1}{2}+\frac{\gamma\Gamma_{12}+\Delta \Omega_{12}}{8|G_{12}|^2}\right)- |\rho_{ge}({\bf r})|^2 - \bar{\nabla}^2_{\perp} \right]\rho_{ge}(\textbf{r}) + \frac{\mathcal{R}}{8G_{12}},
\end{align}
where $G_{12} = \Omega_{12}+\text{i}\Gamma_{12}$, $\bar{t} = -8G_{12}t$ and $\bar{{\bf r}} = 2\sqrt{2}{\bf r}/a$. 
Solving Eq.~\eqref{Eq:DNLSEScaled} for the uniform steady state, we have
\begin{align}\label{Eq:Bistability}
\bigg|\frac{\mathcal{R}}{8G_{12}}\bigg|^2 = |\rho_{ge}|^2\left[\left(\frac{\gamma\Omega_{12}-\Delta \Gamma_{12}}{8|G_{12}|^2}\right)^2+\left(\frac{1}{2}+\frac{\gamma\Gamma_{12}+\Delta \Omega_{12}}{8|G_{12}|^2}-|\rho_{ge}|^2\right)^2\right].
\end{align}
Pattern formation usually occurs for parameters where Eq.~\eqref{Eq:Bistability} exhibits multiple solutions, i.e., bistability (see Fig.~\ref{fig:bistabilityDNLSE}). Taking the derivative of both sides of Eq.~\eqref{Eq:Bistability} with respect to $|\rho_{ge}|^2$, we find bistability is only possible when the necessary (but not sufficient) condition
\begin{align}\label{BistabilityCondition}
4|G_{12}|^2+\gamma\Gamma_{12}+\Delta \Omega_{12} > \sqrt{3}(\gamma\Omega_{12}-\Delta \Gamma_{12}),
\end{align}
is satisfied.

From analysis of Eq.~\eqref{BistabilityCondition}, we find that bistability and patterns emerge in the NLSE for lattice spacings where the collective line shift due to neighboring atom dominates over the other energy scales.
Because of the drastic simplifications of the full numerical model, the NLSE analogy obviously cannot reproduce similar patterns and is not accurate in predicting the parameter values for the onset of bistability or pattern formation. However, it still represents analogous phenomenology, including comparable geometry in the shapes of the patterns (see Fig.~\ref{fig:patternsDNLSE}).
The pattern formation occurs from one or both of the two spatially uniform bistable solutions becoming unstable due to fluctuations in the atomic excitation, which then undergo a four-wave mixing process and lead to the formation of spatially varying stable structures~\cite{Castelli2017}.

%%%%%%%%%%%%%%%%%%%%%%%%%%%%%%%%%%%%%%
\subsection{Patterns with multilevel atomic transition}

\begin{figure}[h]
	\hspace*{-0cm}
	\includegraphics[width=\columnwidth]{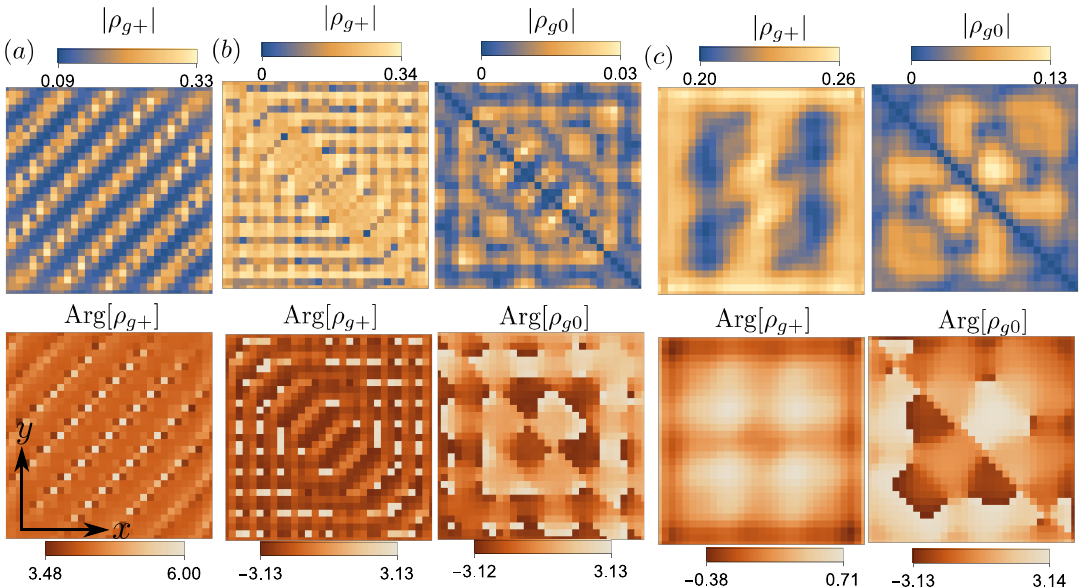}
	\vspace{-0cm}
	\caption{Formation of patterns in the atomic coherence for an $N=961$ atomic array with $a=0.2\lambda$, a $J=0\rightarrow J'=1$ transition and incident plane wave. 
	(a) Striped patterns emerge for an incident field intensity $I/I_{\rm sat}=10$ and laser frequency detuning from the atomic resonance $\Delta/\gamma=0.5$.
	(b,c) A mirror at a distance $d=1.98\lambda$ alters the patterns, resulting in a nonzero out-of-plane ($\rho_{g0}$) polarization component, with (b) a stripe-like phase for $I/I_{\rm sat}=55$ and $\Delta/\gamma=-3.5$, or (c) clumps for $I/I_{\rm sat}=105$ and $\Delta/\gamma=3$.
	}
	\label{fig:patterns4LS}
\end{figure}

We find patterns emerge also when driving a $J=0\rightarrow J'=1$ transition in a square array of atoms.
By evolving Eqs.~\eqref{Eq:SpinEquations} in time, similar checkerboard patterns [Fig.~\ref{fig:patterns4LS}(a)] to the two-level transition dynamics can be found in the $m_{J'}=\pm1$ level coherences, $\rho_{g\pm}$, in the array without a mirror. Both coherences $\rho_{g\pm}$ exhibit the same pattern, but with the $\rho_{g-}$ pattern a reflection of the $\rho_{g+}$ pattern along the line $y=x$ and with a $\pi$ phase change. 
Adding a mirror has more of a substantial effect on the dynamics for the $J=0\rightarrow J'=1$ transition as the dipoles can now rotate out of the atomic plane when driven by the reflected scattered light. This leads to a nonzero $m_{J'}=0$ coherence $\rho_{g0}$ which for fixed atomic positions cannot be present without a mirror.
For certain patterns, $\rho_{g0}$ is much smaller in magnitude than $\rho_{g\pm}$, where both components display similar patterns, e.g., a stripe phase in Fig.~\ref{fig:patterns4LS}(b). However, for other cases, the rotation of the dipoles leads to new patterns not seen for the two-level transition, e.g., Fig.~\ref{fig:patterns4LS}(c), with comparable magnitudes of $\rho_{g\pm}$ and $\rho_{g0}$. 
Similar to the two-level transition dynamics, we find that these patterns get distorted or are sometimes lost when there are fluctuations in the atom positions. 

\begin{figure}
	\hspace*{-0cm}
	\includegraphics[width=\columnwidth]{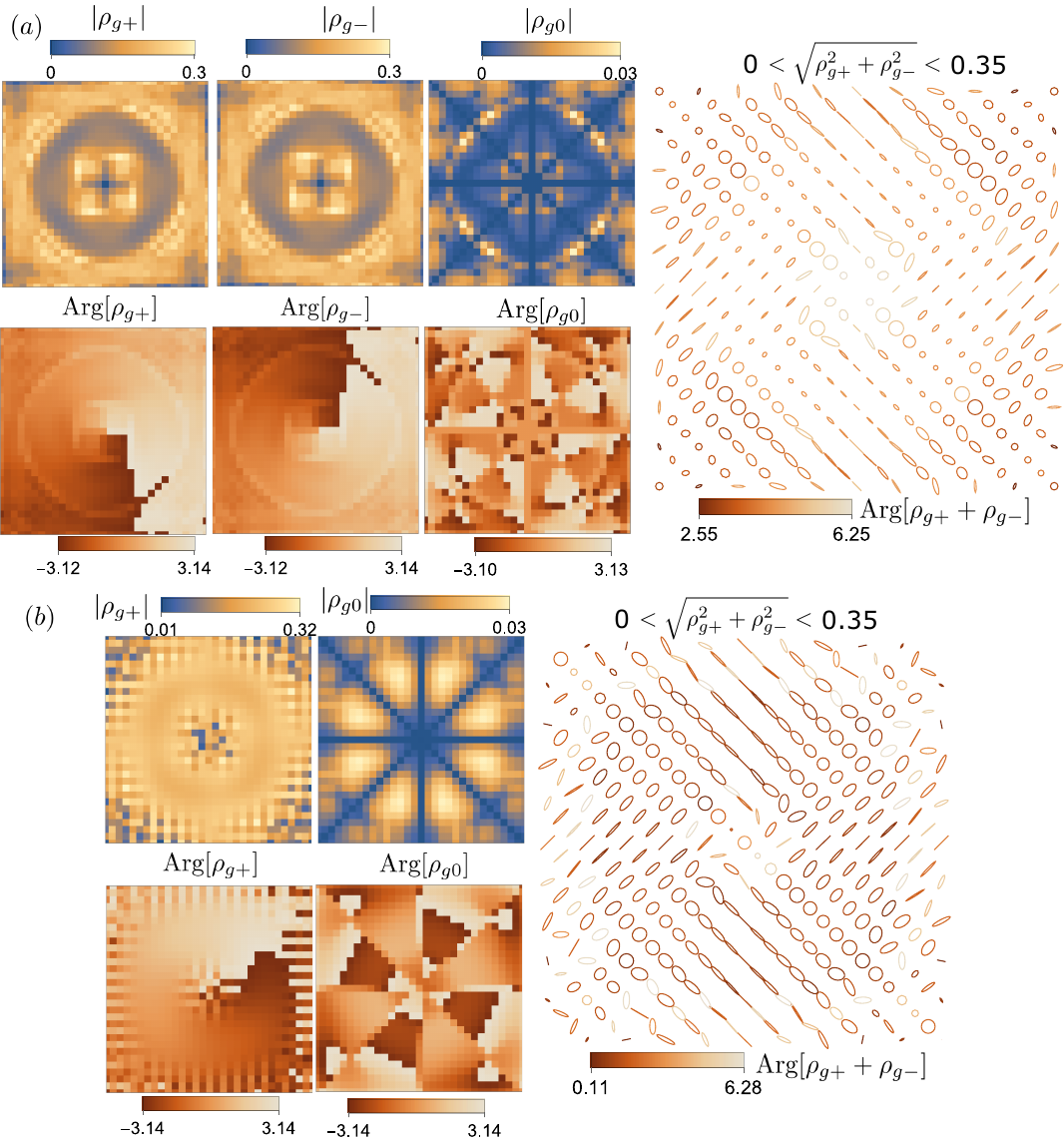}
	\vspace{-0cm}
	\caption{Change in patterns when a field with a singular vortex illuminates the same array as in Fig.~\ref{fig:patterns4LS}. More exotic patterns emerge for an incident field intensity (a) $I/I_{\rm sat}=36$ and laser frequency detuning from the atomic resonance $\Delta/\gamma=0$, and for (b) $I/I_{\rm sat}=36$ and $\Delta/\gamma=3.5$ than compared to a spatially uniform incident field. 
    Polarization ellipses of the in-plane polarization $(\rho_{g+}, \rho_{g-})$ shown for a subset of the atoms give a different view of the patterns, with ellipses scaled so they vanish (take maximal size) for the smallest (largest) magnitude polarization indicated by the numbers at top of each plot.
	}
	\label{fig:patternsVortices}
\end{figure}

Having demonstrated similarities between the cooperatively interacting free-space atom array and nonlinear dynamics in optical Kerr media and ring cavities in the previous section, it is therefore also interesting to consider nonlinear vortex defects. 
Defects can be imprinted by illuminating atoms with a laser beam exhibiting singularities. We consider two Laguerre-Gaussian beams with opposite orbital angular momentum, which results in a singular vortex in the light field~\cite{Donati2016}.
Changing the field profile leads to substantially different patterns, shown in Fig.~\ref{fig:patternsVortices}(a,b). This produces patterns with 8-fold symmetry in the out-of-plane polarization, with phase rotation in all three polarization components.

\section{Conclusions}

The initial interest in pattern formation in nonlinear optics arose from the capability of relatively simple systems to replicate the complex dynamics largely only seen in chemistry and biology. Here, we have demonstrated that an ultrathin layer of laser-driven, dipole-coupled atoms in free space exhibits complex nonlinear phenomenology, resulting in the spontaneous emergence of intricate optical patterns. Our findings reveal that this pattern formation is linked to optical bistability and is inherently due to small fluctuations within the ensemble that are characterized by strong cooperative light-mediated coupling between the atoms. Usually, optical pattern formation necessitates amplification of fluctuations by mirrors or optical cavities. However, the cooperative interactions among free-space atoms in our case eliminate the requirement for such feedback mechanisms. Our work thus presents another example of how strong interactions between regularly spaced atoms can give rise to physical phenomena akin to those associated with optical cavity systems. We have further highlighted the connection between the studied system and the cavities or nonlinear Kerr media by demonstrating that the optical response dynamics can be approximated using a dissipative NLSE, similar to the LLE describing pattern formation in ring cavities. These links between our model system and widely known nonlinear phenomena and pattern formation reveal the possibility of several intriguing effects such as the vortex defects we already briefly addressed, soliton solutions~\cite{Odent2011,Minardi10,Firth96}, and frequency combs~\cite{Castelli2017} with potential technological applications~\cite{Leo2010}. The proposed phenomena could be experimentally observed in light transmission through Mott-insulator states of atoms as studied in Refs.~\cite{Rui2020,Srakaew22}, where short lattice constants are achievable with Sr~\cite{Olmos13,Ballantine22str} and Yb~\cite{Beloy12} atoms.

\section*{Funding}
We acknowledge financial support from the UK EPSRC (Grant No.\ EP/S002952/1).

%\section*{Disclosures}
%The authors declare no conflicts of interest.

%\section*{Data Availability}
%Data used in the publication is available at (DOI TO BE ADDED IN PROOF).


\begin{thebibliography}{10}
\newcommand{\enquote}[1]{``#1''}

\bibitem{Cross1993}
M.~C. Cross and P.~C. Hohenberg, \enquote{Pattern formation outside of
  equilibrium,} {\protect\JournalTitle{Rev. Mod. Phys.}} \textbf{65}, 851--1112
  (1993).

\bibitem{Meinhardt1992}
H.~Meinhardt, \enquote{Pattern formation in biology: a comparison of models and
  experiments,} {\protect\JournalTitle{Reports on Progress in Physics}}
  \textbf{55}, 797 (1992).

\bibitem{Maini1997}
P.~K.~Maini, K.~J.~Painter, and H.~Nguyen Phong~Chau, \enquote{Spatial pattern
  formation in chemical and biological systems,} {\protect\JournalTitle{J.
  Chem. Soc.{,} Faraday Trans.}} \textbf{93}, 3601--3610 (1997).

\bibitem{Nabika2020}
H.~Nabika, M.~Itatani, and I.~Lagzi, \enquote{Pattern formation in
  precipitation reactions: The liesegang phenomenon,}
  {\protect\JournalTitle{Langmuir}} \textbf{36}, 481--497 (2020). PMID:
  31774294.

\bibitem{Turing1952}
A.~M. Turing, \enquote{The chemical basis of morphogenesis,}
  {\protect\JournalTitle{Phil. Trans. R. Soc. Lond. B}} \textbf{237}, 37--72
  (1952).

\bibitem{Fauve1998}
S.~Fauve, \emph{Pattern forming instabilities} (Cambridge University Press,
  1998), p. 387–492, Collection Alea-Saclay: Monographs and Texts in
  Statistical Physics.

\bibitem{Macdonald1992}
R.~Macdonald and H.~Eichler, \enquote{Spontaneous optical pattern formation in
  a nematic liquid crystal with feedback mirror,} {\protect\JournalTitle{Optics
  Communications}} \textbf{89}, 289--295 (1992).

\bibitem{Lugiato1987}
L.~A. Lugiato and R.~Lefever, \enquote{Spatial dissipative structures in
  passive optical systems,} {\protect\JournalTitle{Phys. Rev. Lett.}}
  \textbf{58}, 2209--2211 (1987).

\bibitem{Grynberg1988}
G.~Grynberg, E.~{Le Bihan}, P.~Verkerk, P.~Simoneau, J.~Leite, D.~Bloch, S.~{Le
  Boiteux}, and M.~Ducloy, \enquote{Observation of instabilities due to
  mirrorless four-wave mixing oscillation in sodium,}
  {\protect\JournalTitle{Optics Communications}} \textbf{67}, 363--366 (1988).

\bibitem{Firth1990}
W.~Firth, \enquote{Spatial instabilities in a kerr medium with single feedback
  mirror,} {\protect\JournalTitle{Journal of Modern Optics}} \textbf{37},
  151--153 (1990).

\bibitem{DAlessandro1991}
G.~D'Alessandro and W.~J. Firth, \enquote{Spontaneous hexagon formation in a
  nonlinear optical medium with feedback mirror,} {\protect\JournalTitle{Phys.
  Rev. Lett.}} \textbf{66}, 2597--2600 (1991).

\bibitem{HAELTERMAN1992}
M.~Haelterman, S.~Trillo, and S.~Wabnitz, \enquote{Dissipative modulation
  instability in a nonlinear dispersive ring cavity,}
  {\protect\JournalTitle{Optics Communications}} \textbf{91}, 401--407 (1992).

\bibitem{Ackemann95}
T.~Ackemann, Y.~A. Logvin, A.~Heuer, and W.~Lange, \enquote{Transition between
  positive and negative hexagons in optical pattern formation,}
  {\protect\JournalTitle{Phys. Rev. Lett.}} \textbf{75}, 3450--3453 (1995).

\bibitem{Afanasev1995}
A.~Afanas'ev, Y.~Logvin, A.~Samson, and B.~Samson, \enquote{Spontaneous
  patterns formation in thin film of two-level centers with a wide-aperture
  optical feedback,} {\protect\JournalTitle{Optics Communications}}
  \textbf{115}, 559--562 (1995).

\bibitem{Loiko1996}
N.~Loiko, Y.~Logvin, and A.~Samson, \enquote{Delay instabilities in light
  transmission of thin film with mirror,} {\protect\JournalTitle{Optics
  Communications}} \textbf{124}, 383--391 (1996).

\bibitem{Arecchi1999}
F.~Arecchi, S.~Boccaletti, and P.~Ramazza, \enquote{Pattern formation and
  competition in nonlinear optics,} {\protect\JournalTitle{Physics Reports}}
  \textbf{318}, 1--83 (1999).

\bibitem{Schapers00}
B.~Sch\"apers, M.~Feldmann, T.~Ackemann, and W.~Lange, \enquote{Interaction of
  localized structures in an optical pattern-forming system,}
  {\protect\JournalTitle{Phys. Rev. Lett.}} \textbf{85}, 748--751 (2000).

\bibitem{Castelli2017}
F.~Castelli, M.~Brambilla, A.~Gatti, F.~Prati, and L.~A. Lugiato, \enquote{The
  lle, pattern formation and a novel coherent source,}
  {\protect\JournalTitle{The European Physical Journal D}} \textbf{71},
  1434--6079 (2017).

\bibitem{Scroggie94}
A.~Scroggie, W.~Firth, G.~McDonald, M.~Tlidi, R.~Lefever, and L.~Lugiato,
  \enquote{Pattern formation in a passive kerr cavity,}
  {\protect\JournalTitle{Chaos, Solitons \& Fractals}} \textbf{4}, 1323--1354
  (1994). Special Issue: Nonlinear Optical Structures, Patterns, Chaos.

\bibitem{Domokos2002}
P.~Domokos and H.~Ritsch, \enquote{Collective cooling and self-organization of
  atoms in a cavity,} {\protect\JournalTitle{Phys. Rev. Lett.}} \textbf{89},
  253003 (2002).

\bibitem{Black03}
A.~T. Black, H.~W. Chan, and V.~Vuleti\ifmmode~\acute{c}\else \'{c}\fi{},
  \enquote{Observation of collective friction forces due to spatial
  self-organization of atoms: From {Rayleigh} to {Bragg} scattering,}
  {\protect\JournalTitle{Phys. Rev. Lett.}} \textbf{91}, 203001 (2003).

\bibitem{Asboth2005}
J.~K. Asb{\'{o}}th, P.~Domokos, H.~Ritsch, and A.~Vukics,
  \enquote{{Self-organization of atoms in a cavity field: Threshold,
  bistability, and scaling laws},} {\protect\JournalTitle{Phys. Rev. A}}
  \textbf{72}, 053417 (2005).

\bibitem{Lee14}
M.~D. Lee and J.~Ruostekoski, \enquote{Classical stochastic measurement
  trajectories: Bosonic atomic gases in an optical cavity and quantum
  measurement backaction,} {\protect\JournalTitle{Phys. Rev. A}} \textbf{90},
  023628 (2014).

\bibitem{Baumann2010}
K.~Baumann, C.~Guerlin, F.~Brennecke, and T.~Esslinger, \enquote{{Dicke quantum
  phase transition with a superfluid gas in an optical cavity},}
  {\protect\JournalTitle{Nature}} \textbf{464}, 1301--1306 (2010).

\bibitem{Caballero-Benitez2015}
S.~F. Caballero-Benitez and I.~B. Mekhov, \enquote{Quantum optical lattices for
  emergent many-body phases of ultracold atoms,} {\protect\JournalTitle{Phys.
  Rev. Lett.}} \textbf{115}, 243604 (2015).

\bibitem{Vaidya18}
V.~D. Vaidya, Y.~Guo, R.~M. Kroeze, K.~E. Ballantine, A.~J. Koll\'ar,
  J.~Keeling, and B.~L. Lev, \enquote{Tunable-range, photon-mediated atomic
  interactions in multimode cavity {QED},} {\protect\JournalTitle{Phys. Rev.
  X}} \textbf{8}, 011002 (2018).

\bibitem{Labeyrie2014}
G.~Labeyrie, E.~Tesio, P.~M. Gomes, G.-L. Oppo, W.~J. Firth, G.~R.~M. Robb,
  A.~S. Arnold, R.~Kaiser, and T.~Ackemann, \enquote{{Optomechanical
  self-structuring in a cold atomic gas},} {\protect\JournalTitle{Nat.
  Photonics}} \textbf{8}, 321--325 (2014).

\bibitem{Baio21}
G.~Baio, G.~R.~M. Robb, A.~M. Yao, G.-L. Oppo, and T.~Ackemann,
  \enquote{Multiple self-organized phases and spatial solitons in cold atoms
  mediated by optical feedback,} {\protect\JournalTitle{Phys. Rev. Lett.}}
  \textbf{126}, 203201 (2021).

\bibitem{meystre1998}
P.~Meystre and M.~Sargent, \emph{Elements of Quantum Optics} (Springer Berlin
  Heidelberg, 1998).

\bibitem{Odent2011}
V.~Odent, M.~Taki, and E.~Louvergneaux, \enquote{Experimental evidence of
  dissipative spatial solitons in an optical passive kerr cavity,}
  {\protect\JournalTitle{New J. Phys.}} \textbf{13}, 113026 (2011).

\bibitem{Minardi10}
S.~Minardi, F.~Eilenberger, Y.~V. Kartashov, A.~Szameit, U.~R\"opke,
  J.~Kobelke, K.~Schuster, H.~Bartelt, S.~Nolte, L.~Torner, F.~Lederer,
  A.~T\"unnermann, and T.~Pertsch, \enquote{Three-dimensional light bullets in
  arrays of waveguides,} {\protect\JournalTitle{Phys. Rev. Lett.}}
  \textbf{105}, 263901 (2010).

\bibitem{Firth96}
W.~J. Firth and A.~J. Scroggie, \enquote{Optical bullet holes: Robust
  controllable localized states of a nonlinear cavity,}
  {\protect\JournalTitle{Phys. Rev. Lett.}} \textbf{76}, 1623--1626 (1996).

\bibitem{Azam2022}
P.~Azam, A.~Griffin, S.~Nazarenko, and R.~Kaiser, \enquote{Vortex creation,
  annihilation, and nonlinear dynamics in atomic vapors,}
  {\protect\JournalTitle{Phys. Rev. A}} \textbf{105}, 043510 (2022).

\bibitem{Rui2020}
J.~Rui, D.~Wei, A.~Rubio-Abadal, S.~Hollerith, J.~Zeiher, D.~M. Stamper-Kurn,
  C.~Gross, and I.~Bloch, \enquote{{A subradiant optical mirror formed by a
  single structured atomic layer},} {\protect\JournalTitle{Nature}}
  \textbf{583}, 369--374 (2020).

\bibitem{Srakaew22}
K.~Srakaew, P.~Weckesser, S.~Hollerith, D.~Wei, D.~Adler, I.~Bloch, and
  J.~Zeiher, \enquote{A subwavelength atomic array switched by a single rydberg
  atom,} {\protect\JournalTitle{Nature Physics}} \textbf{19}, 714--719 (2023).

\bibitem{Jenkins2012a}
S.~D. Jenkins and J.~Ruostekoski, \enquote{Controlled manipulation of light by
  cooperative response of atoms in an optical lattice,}
  {\protect\JournalTitle{Phys. Rev. A}} \textbf{86}, 031602 (2012).

\bibitem{Ruostekoski23}
J.~Ruostekoski, \enquote{Cooperative quantum-optical planar arrays of atoms,}
  {\protect\JournalTitle{Phys. Rev. A}} \textbf{108}, 030101 (2023).

\bibitem{Yu14}
N.~Yu and F.~Capasso, \enquote{Flat optics with designer metasurfaces,}
  {\protect\JournalTitle{Nature Materials}} \textbf{13}, 139--150 (2014).

\bibitem{Parmee2018}
C.~D. Parmee and N.~R. Cooper, \enquote{Phases of driven two-level systems with
  nonlocal dissipation,} {\protect\JournalTitle{Phys. Rev. A}} \textbf{97},
  053616 (2018).

\bibitem{williamson2020b}
L.~A. Williamson, M.~O. Borgh, and J.~Ruostekoski, \enquote{Superatom picture
  of collective nonclassical light emission and dipole blockade in atom
  arrays,} {\protect\JournalTitle{Phys. Rev. Lett.}} \textbf{125}, 073602
  (2020).

\bibitem{Cidrim20}
A.~Cidrim, T.~S. {do Espirito Santo}, J.~Schachenmayer, R.~Kaiser, and
  R.~Bachelard, \enquote{Photon blockade with ground-state neutral atoms,}
  {\protect\JournalTitle{Phys. Rev. Lett.}} \textbf{125}, 073601 (2020).

\bibitem{Parmee2020}
C.~D. Parmee and J.~Ruostekoski, \enquote{{Signatures of optical phase
  transitions in superradiant and subradiant atomic arrays},}
  {\protect\JournalTitle{Commun. Phys.}} \textbf{3}, 205 (2020).

\bibitem{Parmee2021}
C.~D. Parmee and J.~Ruostekoski, \enquote{Bistable optical transmission through
  arrays of atoms in free space,} {\protect\JournalTitle{Phys. Rev. A}}
  \textbf{103}, 033706 (2021).

\bibitem{Bettles2020}
R.~J. Bettles, M.~D. Lee, S.~A. Gardiner, and J.~Ruostekoski, \enquote{{Quantum
  and nonlinear effects in light transmitted through planar atomic arrays},}
  {\protect\JournalTitle{Commun. Phys.}} \textbf{3}, 141 (2020).

\bibitem{Zhang2022}
L.~Zhang, V.~Walther, K.~M{\o{}}lmer, and T.~Pohl, \enquote{Photon-photon
  interactions in {R}ydberg-atom arrays,} {\protect\JournalTitle{{Quantum}}}
  \textbf{6}, 674 (2022).

\bibitem{Ferioli21}
G.~Ferioli, A.~Glicenstein, L.~Henriet, I.~Ferrier-Barbut, and A.~Browaeys,
  \enquote{Storage and release of subradiant excitations in a dense atomic
  cloud,} {\protect\JournalTitle{Phys. Rev. X}} \textbf{11}, 021031 (2021).

\bibitem{Holzinger21}
R.~Holzinger, M.~Moreno-Cardoner, and H.~Ritsch, \enquote{Nanoscale continuous
  quantum light sources based on driven dipole emitter arrays,}
  {\protect\JournalTitle{Applied Physics Letters}} \textbf{119}, 024002 (2021).

\bibitem{Rusconi2021}
C.~C. Rusconi, T.~Shi, and J.~I. Cirac, \enquote{Exploiting the photonic
  nonlinearity of free-space subwavelength arrays of atoms,}
  {\protect\JournalTitle{Phys. Rev. A}} \textbf{104}, 033718 (2021).

\bibitem{Moreno2021}
M.~Moreno-Cardoner, D.~Goncalves, and D.~E. Chang, \enquote{Quantum nonlinear
  optics based on two-dimensional {Rydberg} atom arrays,}
  {\protect\JournalTitle{Phys. Rev. Lett.}} \textbf{127}, 263602 (2021).

\bibitem{Parmee22b}
C.~D. Parmee, K.~E. Ballantine, and J.~Ruostekoski, \enquote{Spontaneous
  symmetry breaking in frustrated triangular atom arrays due to cooperative
  light scattering,} {\protect\JournalTitle{Phys. Rev. Res.}} \textbf{4},
  043039 (2022).

\bibitem{Pedersen23}
S.~P. Pedersen, L.~Zhang, and T.~Pohl, \enquote{Quantum nonlinear metasurfaces
  from dual arrays of ultracold atoms,} {\protect\JournalTitle{Phys. Rev.
  Res.}} \textbf{5}, L012047 (2023).

\bibitem{Rubies-Bigorda23}
O.~Rubies-Bigorda, S.~Ostermann, and S.~F. Yelin, \enquote{Characterizing
  superradiant dynamics in atomic arrays via a cumulant expansion approach,}
  {\protect\JournalTitle{Phys. Rev. Res.}} \textbf{5}, 013091 (2023).

\bibitem{Robicheaux23}
F.~Robicheaux and D.~A. Suresh, \enquote{Intensity effects of light coupling to
  one- or two-atom arrays of infinite extent,} {\protect\JournalTitle{Phys.
  Rev. A}} \textbf{108}, 013711 (2023).

\bibitem{CohenT}
C.~{Cohen-Tannaudji}, J.~{Dupont-Roc}, and G.~Grynberg, \emph{Photons and
  Atoms: Introduction to Quantum Electrodynamics} (John Wiley \& Sons, New
  York, 1989).

\bibitem{Ruostekoski1997a}
J.~Ruostekoski and J.~Javanainen, \enquote{Quantum field theory of cooperative
  atom response: Low light intensity,} {\protect\JournalTitle{Phys. Rev. A}}
  \textbf{55}, 513--526 (1997).

\bibitem{Lee16}
M.~D. Lee, S.~D. Jenkins, and J.~Ruostekoski, \enquote{Stochastic methods for
  light propagation and recurrent scattering in saturated and nonsaturated
  atomic ensembles,} {\protect\JournalTitle{Phys. Rev. A}} \textbf{93}, 063803
  (2016).

\bibitem{Jackson}
J.~D. Jackson, \emph{Classical Electrodynamics} (Wiley, New York, 1999), 3rd
  ed.

\bibitem{Javanainen2019}
J.~Javanainen and R.~Rajapakse, \enquote{Light propagation in systems involving
  two-dimensional atomic lattices,} {\protect\JournalTitle{Phys. Rev. A}}
  \textbf{100}, 013616 (2019).

\bibitem{Donati2016}
S.~Donati, L.~Dominici, G.~Dagvadorj, D.~Ballarini, M.~{De Giorgi}, A.~Bramati,
  G.~Gigli, Y.~G. Rubo, M.~H. Szyma{\'{n}}ska, and D.~Sanvitto, \enquote{{Twist
  of generalized skyrmions and spin vortices in a polariton superfluid},}
  {\protect\JournalTitle{Proceedings of the National Academy of Sciences}}
  \textbf{113}, 14926--14931 (2016).

\bibitem{Leo2010}
F.~Leo, S.~Coen, P.~Kockaert, and et~al., \enquote{Temporal cavity solitons in
  one-dimensional kerr media as bits in an all-optical buffer.}
  {\protect\JournalTitle{Nature Photon}} \textbf{4}, 471--476 (2010).

\bibitem{Olmos13}
B.~Olmos, D.~Yu, Y.~Singh, F.~Schreck, K.~Bongs, and I.~Lesanovsky,
  \enquote{Long-range interacting many-body systems with alkaline-earth-metal
  atoms,} {\protect\JournalTitle{Phys. Rev. Lett.}} \textbf{110}, 143602
  (2013).

\bibitem{Ballantine22str}
K.~E. Ballantine, D.~Wilkowski, and J.~Ruostekoski, \enquote{Optical magnetism
  and wavefront control by arrays of strontium atoms,}
  {\protect\JournalTitle{Phys. Rev. Res.}} \textbf{4}, 033242 (2022).

\bibitem{Beloy12}
K.~Beloy, J.~A. Sherman, N.~D. Lemke, N.~Hinkley, C.~W. Oates, and A.~D.
  Ludlow, \enquote{Determination of the $5d6s$ ${}^{3}{D}_{1}$ state lifetime
  and blackbody-radiation clock shift in yb,} {\protect\JournalTitle{Phys. Rev.
  A}} \textbf{86}, 051404 (2012).

\end{thebibliography}
\end{document}